\documentclass[journal,comsoc]{IEEEtran}
\usepackage[T1]{fontenc}

\usepackage{graphicx}
\usepackage{textcomp}
\usepackage{xcolor}
\usepackage{color}
\usepackage{epstopdf}
\usepackage{amsthm}
\usepackage[ruled]{algorithm2e}

\newcommand{\RNum}[1]{\uppercase\expandafter{\romannumeral #1\relax}}
\usepackage{makecell}
\usepackage{multirow}
\usepackage{booktabs}
\usepackage{xpatch}
\usepackage{footnote}
\usepackage[caption=false]{subfig}
\makeatletter
\AtBeginDocument{\xpatchcmd{\@thm}{\thm@headpunct{.}}{\thm@headpunct{:}}{}{}}
\makeatother
\usepackage{soul}
\soulregister\cite7 
\soulregister\citep7 
\soulregister\citet7 
\soulregister\ref7 
\soulregister\eqref7 
\soulregister\pageref7 

\usepackage{tikz}

\ifCLASSINFOpdf
\else
\fi

\usepackage{amsmath}
\usepackage[cmintegrals]{newtxmath}

\begin{document}

\title{Zero Trust Architecture for 6G Security}

\author{Xu~Chen, Wei~Feng, Ning~Ge, and Yan~Zhang
\thanks{X. Chen, W. Feng and N. Ge are with the Department
of Electronic Engineering, Tsinghua University, Beijing 100084, China. (e-mail: chenxu18@mails.tsinghua.edu.cn, fengwei@tsinghua.edu.cn, gening@tsinghua.edu.cn).}
\thanks{Y. Zhang is with the Department of Informatics, University of Oslo, 0316 Oslo, Norway (e-mail: yanzhang@ieee.org).}
}

\maketitle

\begin{abstract}
The upcoming sixth generation (6G) network is envisioned to be more open and heterogeneous than earlier generations. This challenges conventional security architectures, which typically rely on the construction of a security perimeter at network boundaries. In this article, we propose a software-defined zero trust architecture (ZTA) for 6G networks, which is promising for establishing an elastic and scalable security regime. This architecture achieves secure access control through adaptive collaborations among the involved control domains, and can effectively prevent malicious access behaviors such as distributed denial of service (DDoS) attacks, malware spread, and zero-day exploits. We also introduce key design aspects of this architecture and show the simulation results of a case study, which shows the effectiveness and robustness of ZTA for 6G. Furthermore, we discuss open issues to further promote this new architecture.  
\end{abstract}

\IEEEpeerreviewmaketitle

\section{Introduction}
Currently, the sixth-generation (6G) network is being extensively studied. One key challenge is on the security issue. 6G is envisioned to provide globally seamless connections to trillions of humans, machines, and things \cite{Fang}. Many types of such devices have weak security capabilities, and can be easily compromised to engage in malicious activities. Meanwhile, the wide application of open-source software technologies also introduces security risks caused by software vulnerabilities. Worse yet, 6G is an open and space-air-ground integrated network, existing perimeter-based security solutions such as firewalls and intrusion detection systems (IDSs), may lead to poor defense effects \cite{csm}. It becomes necessary to build more elastic security architectures to satisfy the requirements of 6G.

To bridge this gap, a collaborative zero trust architecture (ZTA) for 6G is developed in this article. This architecture considers the network control domains as communities and harmonizes them to implement sophisticated access control to protect their internal network entities. In particular, a decentralized network access control regime is established through adaptive collaborations among the involved control domains. Considering the limitations of communities in terms of available resources, blockchain-based third-party security services are introduced to provide computing power and global information support. By simulating the worm spreading process in communities, this architecture can effectively defend against the spread of computer viruses, distributed denial of service (DDoS) attacks, and zero-day exploits through dynamic access control of the cross-community access behaviors of user equipment (UE). To facilitate the implementation of this ZTA, we present open issues for future research, such as system efficiency, mobility management, trust evaluation and residual security risks. 
\section{Challenges in 6G Security}
\subsection{Attack model}
UEs and network entities in 6G networks are vulnerable to many types of threats. The authentication mechanisms in 6G may fail to prevent legitimate but compromised UEs from engaging in malicious activities \cite{survey}. We summarize malicious behaviors in the network and transport layers, including DDoS attacks, malware, and zero-day exploits.
\begin{itemize}
	\item \textit{DDoS attacks}. DDoS attacks are expected to be more serious for 6G. Many insecure internet of things (IoT) devices may provide attackers with large-scale malicious traffic. However, widely used edge servers have weaker mitigation capabilities against DDoS attacks. UEs are also vulnerable to DDoS attacks.
	\item \textit{Malware}. Malware is a common software tool for a large number of malicious behaviors. There are many types of malware, such as computer viruses, worms, Trojan horses, ransomware and rootkits. Malware is usually designed to steal data and disrupt the operation of software systems, thereby causing damage to 6G networks.
	\item \textit{Zero-day exploits}. Zero-day exploits are another type of software-based malicious behavior. Software vulnerability is often difficult to avoid. Because 6G networks widely adopt open-source software to implement network entity functions, attackers can exploit zero-day vulnerabilities to adversely affect network functions.
\end{itemize}
\begin{figure*}[t]
	\centerline{\includegraphics[width=0.7\textwidth]{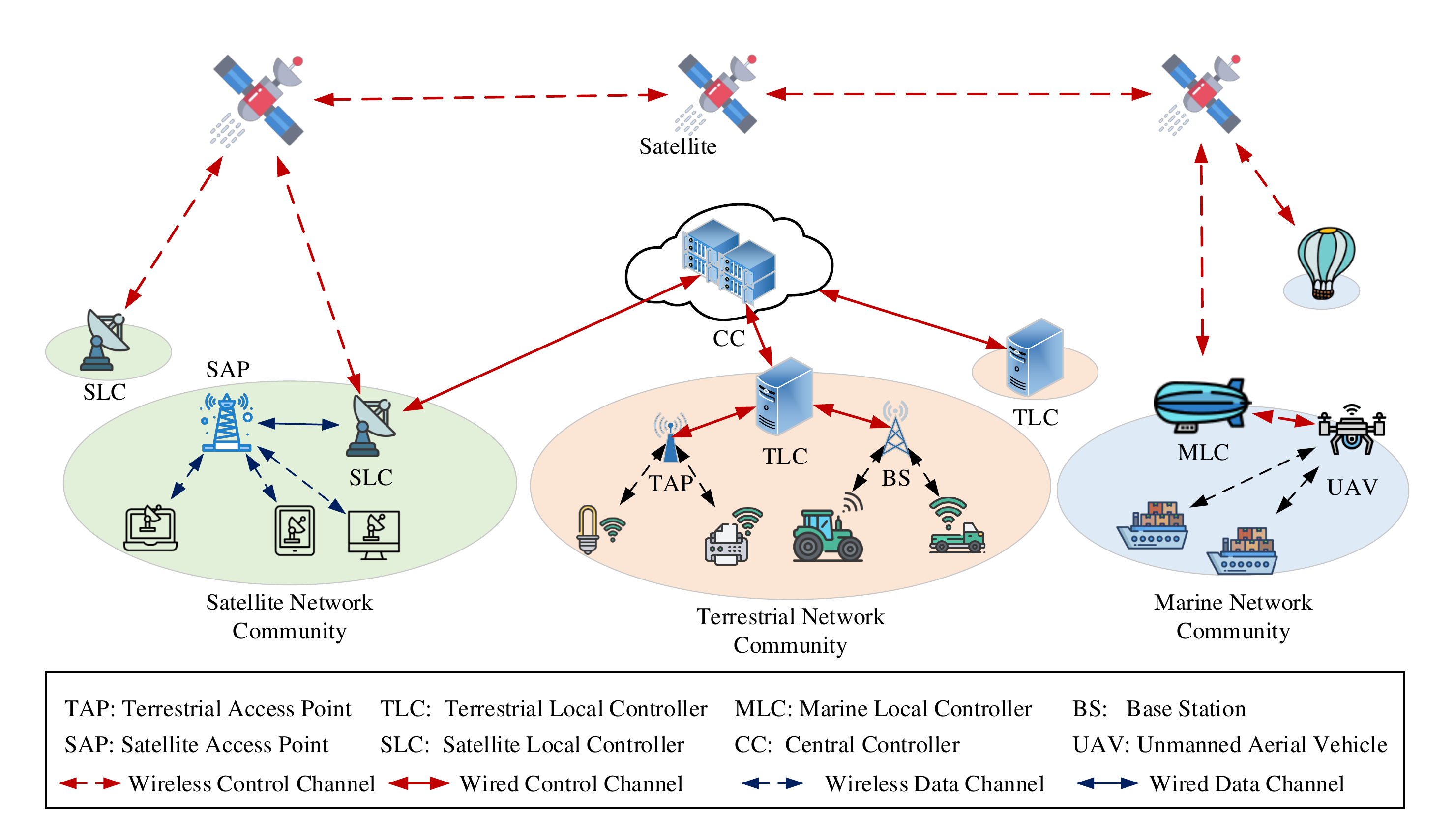}}
	\caption{6G network architecture. The network managed by each local SDN controller constitutes a community.}
	\label{fig1}
\end{figure*}
\subsection{Security challenges}
We summarize new security challenges for 6G as follows.
\begin{itemize}
	\item Ultra large scale. As an ultra-large-scale network, 6G places an extremely high demand on the scalability of the security architecture. High-Complexity architectures may not be cost effective or even inapplicable at this scale.
	\item Heterogeneity. The heterogeneous networks in 6G will probably be managed by multiple operators with different management architectures and signaling systems. Collaboration among different control domains with various management modes is essential for the security architectures in 6G. A centralized architecture will probably fail to meet these requirements.	
	\item Diversity of UEs. The main UE types for 6G networks are machines and devices. The resources and capabilities of these UEs vary widely. The security architecture should be lightweight. Resource-consuming operations, such as end-to-end encryption, cannot be used as a secure basis for security architectures. 	
\end{itemize}
\subsection{ZTA and its limitations in 6G}
ZTA was proposed to protect digital resources in local networks in a dynamic way, the core idea of which is ``never trust and always verify." \cite{BeyondZT, book2017} When traditional network boundaries are blurred, it becomes quite suitable for dynamic access control to avoid security problems. Several enterprises have conducted relevant technical research and engineering practices. Gartner proposed the adaptive security architecture (ASA) in 2014 and gradually developed it into the zero trust network access (ZTNA). Google has developed the BeyondCorp security model based on its own internal network security governance requirements since 2014 \cite{BeyondCorp}. A report surveyed more than 400 cybersecurity decision makers and revealed that $72\%$ of organizations plan to assess or implement zero trust capabilities before 2021. Gartner predicted that $60\%$ of these enterprises will migrate to ZTNA by 2023 \cite{progress2020}.

However, the existing ZTAs cannot overcome the security challenges faced by 6G networks. First, current ZTAs implement fine-grained access control strategies to protect all data resources and computing services \cite{NIST}. This characteristic cannot cope with the challenges introduced by the ultra large-scale of 6G networks. Second, existing ZTAs are mainly designed for a single network domain with a logically centralized controller. They cannot be applied to heterogeneous 6G networks with decentralized management architectures. Third, end-to-end encryption is mandatory for existing ZTAs \cite{book2017}. Massive IoT terminals in 6G networks cannot meet this requirement owing to resource limitations. Lastly, existing ZTAs aim to prevent data breaches and limit internal lateral movement. They cannot satisfy the security requirements of 6G, such as the prevention of flooding attacks. 
Although the newly emerged software-defined perimeter (SDP) technology \cite{SDP} has extended the concept of ZTA to the network and transport layers, there are still few considerations of other challenges. The tailoring of ZTA for 6G large-scale network security has become an important issue.
\section{Zero Trust Architecture for 6G}
We introduce the key design aspects of ZTA for 6G, which include the distributed security architecture, decentralized identity management, and trust system.
\subsection{Distributed security architecture}
As 6G is an integrated network operated by many different operators, the SDN-based control architecture should be distributed and adaptive. For example, in remote areas, distributed control architectures are suitable for efficiency considerations, whereas in urban areas, a hierarchical control architecture is required for scalability \cite{You}. To ensure its deployability, the ZTA for 6G must be adapted to the hybrid control architecture.

\begin{figure}[t]
	\centerline{\includegraphics[width=0.48\textwidth]{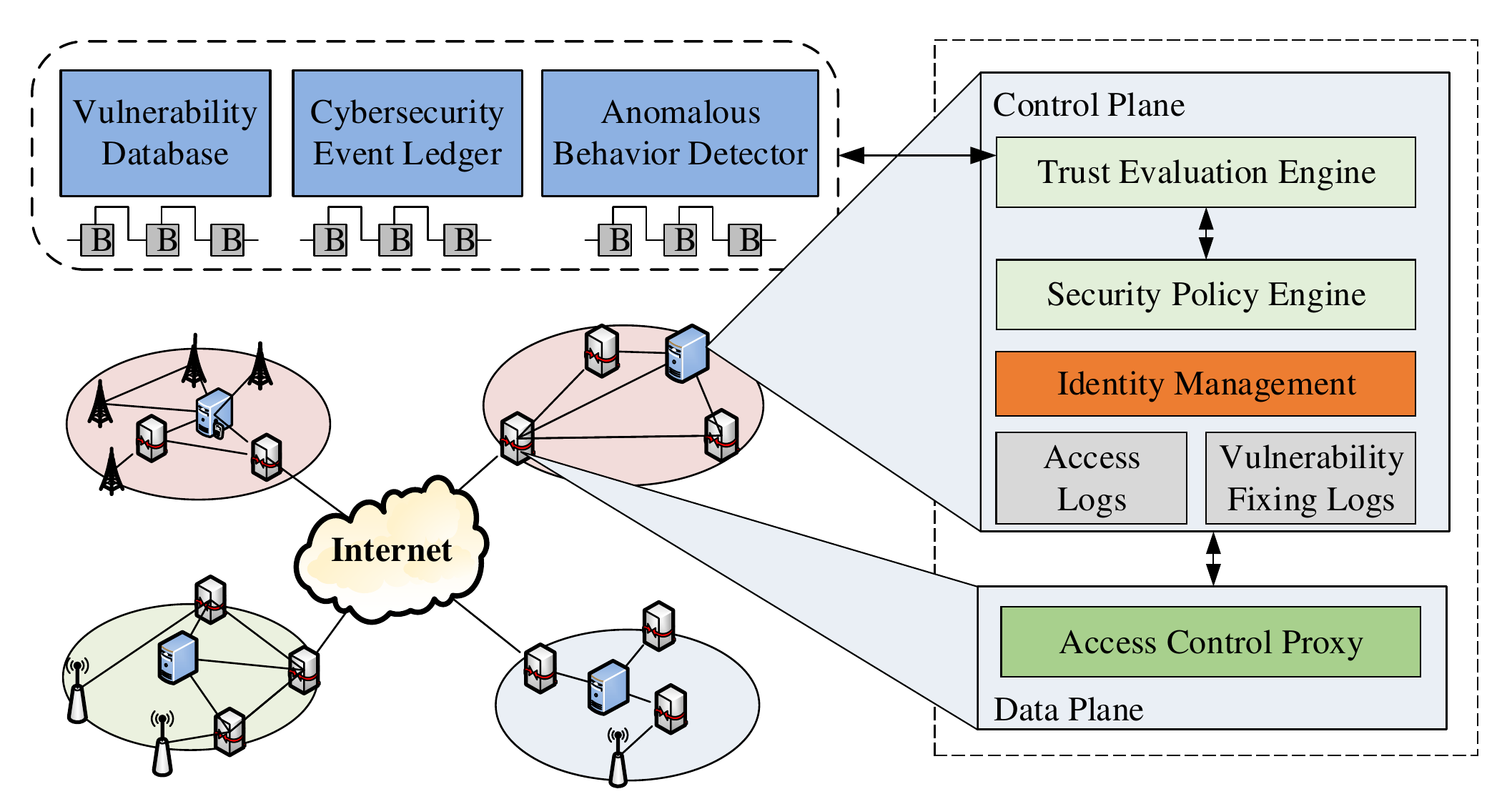}}
	\caption{Access control framework based on communities in ZTA.}
	\label{fig2}
\end{figure}

As shown in Fig.~\ref{fig1}, a community is a subnetwork managed by a local SDN controller in 6G. The security function of communities in ZTA is to perform access control at border switches to protect the internal equipment from external threats. The access control framework for the communities is shown in Fig. \ref{fig2}. The access control proxy on the data plane submits visitors' access requests to the control plane for disposal, and then implements the returned access control decisions. The control plane includes three modules, an identity management module, a security policy engine (SPE), and trust evaluation engine (TEE). The SPE completes the formulation of access control strategies for specific access requests with the support of TEE. The TEE evaluates the trustworthiness of visitors based on third-party security assessment results. The identity management module is responsible for the identity verification and certificate approval of all UEs in the community. In addition, there are two databases to store the vulnerability fixing logs and access logs of UEs, respectively.

To overcome the resource and information limitations of the communities, we introduce third-party security services (TPSSs) based on public blockchains, including the vulnerability database (VDB), cybersecurity event ledger (CEL) and anomalous behavior detector (ABD). The VDB provides risk assessment services to the TEE of the accessed community by analyzing recent vulnerability releases and vulnerability fixing logs related to the guest UE. The risk levels of vulnerable UEs are determined by the type and harmfulness of unfixed vulnerabilities. The CEL checks if the guest UE has communicated with a virus-infected UE and if it has accessed the victims during recent attack events by searching the records obtained through victim reports and network forensics. ABD performs behavior analysis using artificial intelligence techniques and explores potential security risks based on the guest UE's recent access behaviors. The behavior analysis request is usually initiated by the TEE of the accessed community, and the guest UE's home community provides historical access logs on demand. In applications, the blockchain technology enables data integrity and availability for the TPSSs.

When a UE initiates an access request, the home community evaluates and blocks the request if the UE is potentially malicious. After self-evaluation, the UE's digital identity and related security certifications are submitted to the accessed community for evaluation. In the accessed community, trust evaluation is completed in collaboration with TPSSs. When necessary, the SPE sends assessment requests to the TPSS after accepting an access request from the data plane. The TPSS assesses the security risk and returns the results to the TEE. The TEE calculates the trust value based on the feedback from the TPSS and relays the results to the SPE. The SPE makes access control decisions according to the community's security control policies and trust value received from the TEE. This community-based security architecture can effectively adapt to the large scale and heterogeneity of 6G networks.
\subsection{Decentralized identity management}
Identity is the most important infrastructure in ZTAs. In 6G, it is very difficult to establish and maintain a unified identity system due to more diversified network equipment suppliers, network operators and heterogeneous network structures. The dynamically allocated IP addresses cannot be used as identities, including IPv6, because they can be easily spoofed. Therefore, identity management is one of the primary challenges for ZTA in 6G. Traditional identity authentication schemes based on data certificates cannot satisfy the requirements of access control in 6G. In centralized identity management schemes, a unified certification authority (CA) issues certificates for all requesting UEs, which restricts the scalability of the entire system. If multiple CAs are employed, the problem of mutual trust cannot be resolved. In totally distributed identity management schemes, if the service provider (e.g., the web servers) issues a digital certificate to each potential visitor, a UE may retain too many digital certificates. 
 \begin{figure}[t]
	\centerline{\includegraphics[width=0.45\textwidth]{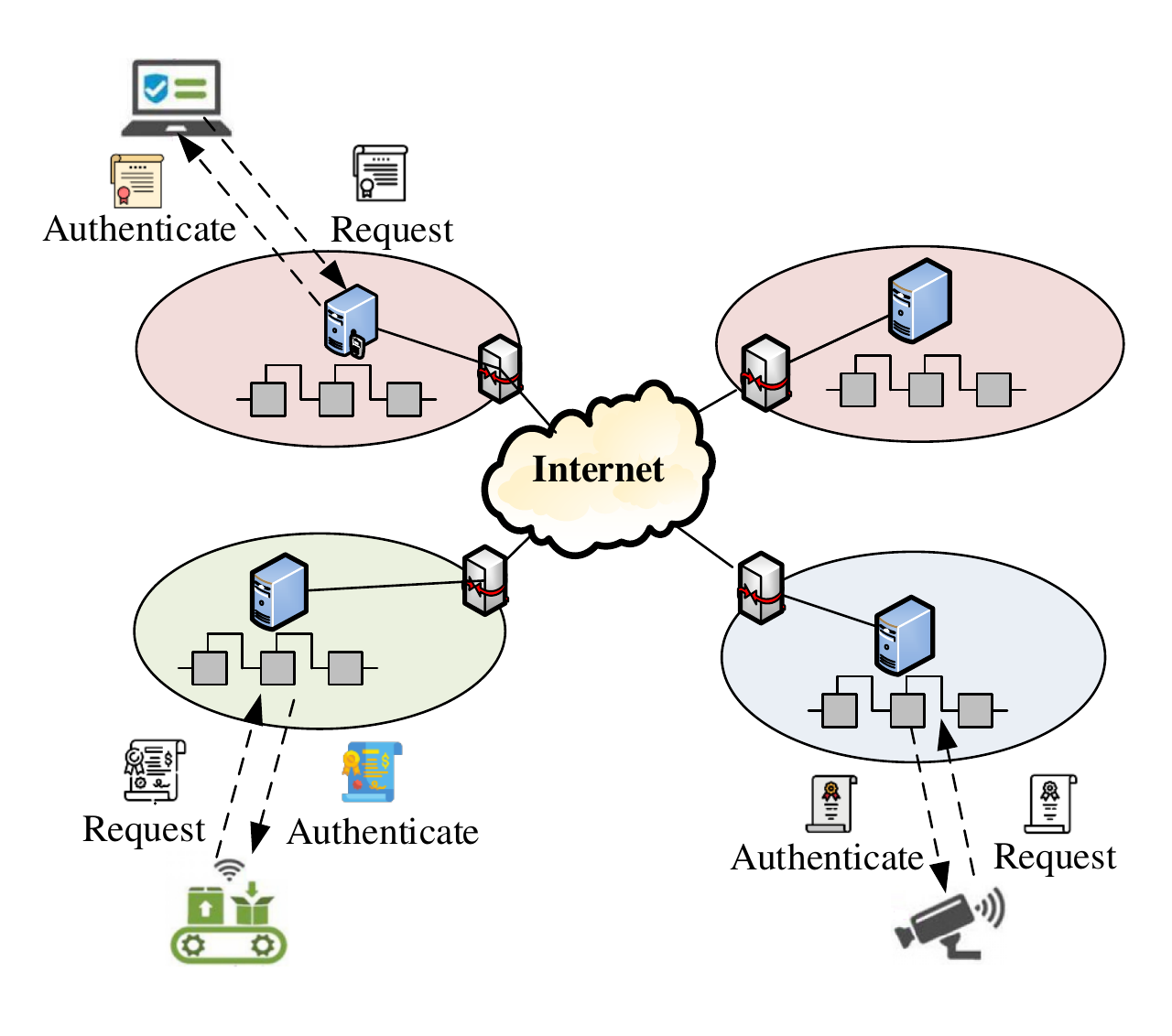}}
	\caption{Community-based decentralized identity authentication.}
	\label{fig3}
\end{figure}

In fact, a unique identity across the entire network is not only unnecessary but also leads to privacy issues. Digital twin technologies have provided us with a distributed identity mechanism based on blockchain \cite{DT}. However, such a unified identity system requires additional privacy protection measures for user data. The complex calculations involved may limit the efficiency and scalability of access control in 6G enabled IoTs. To this end, we propose a decentralized identity management scheme based on digital certificates. In this scheme, the certificate can be generated by either the UE or the home communities. The local controller of each community is the only CA responsible for the authentication of digital certificates. The community-based identity management scheme is illustrated in Fig. \ref{fig3}.
\begin{itemize}
	\item Certificate generation. The digital certificate can be autonomously generated by the UEs. The format of this certificate can be customized by the community. The certificate should specify the proof of identity, UE type, operating system version, and other related information. After generation, certificates should be packaged and encrypted, and then submitted to the local controller for review. The community only needs to file and manage authorized certificates. This scheme can effectively alleviate controller workload. Considering the resource limitations of UEs, certificates can also be generated by the home community.
	\item Certificate registration. The community controller decrypts the submitted certificate packets and verifies relevant information. If successful, a unique certificate number, a validity period and other restrictions are assigned. After signing with the controller's private key, the certificate is returned to the applicant UE as a legal identifier. The registration information of certificates can be stored centrally on the controller or distributed on the community’s private blockchain according to the security requirements.
	\item Certificate updating. If a UE needs to update its digital certificate, it updates the relevant content of the certificate, such as the public key, and performs the certificate generation and registration process again. The controller can use the original certificate ID. It suffices to review, resign and return it to its holder.
	\item Certificate revocation. When necessary, the digital identity can be revoked. During certificate revocations, the controller cancels the certificate ID from the identity database. The controller can also implement a certificate revocation by setting a short validity period.
\end{itemize}

Based on this identity authentication scheme, UE identification can be achieved through hierarchical retrieval by introducing community identifiers (IDs). Specifically, an integrated UE identity consists of three parts: an autonomous system number (ASN), community ID, and certificate ID. The ASN is unique within the entire 6G network. The community ID needs to be unique within the AS to which it belongs, and the certificate ID of a UE needs to be unique within the community in which it resides. The ASN and community ID should be attached as identification elements in cross-domain access. This design not only meets the access control requirements of diversified UEs in 6G but also increases the elasticity and scalability of the identity management scheme. 
\subsection{Trust system}
\begin{figure}[t]
	\centerline{\includegraphics[width=0.5\textwidth]{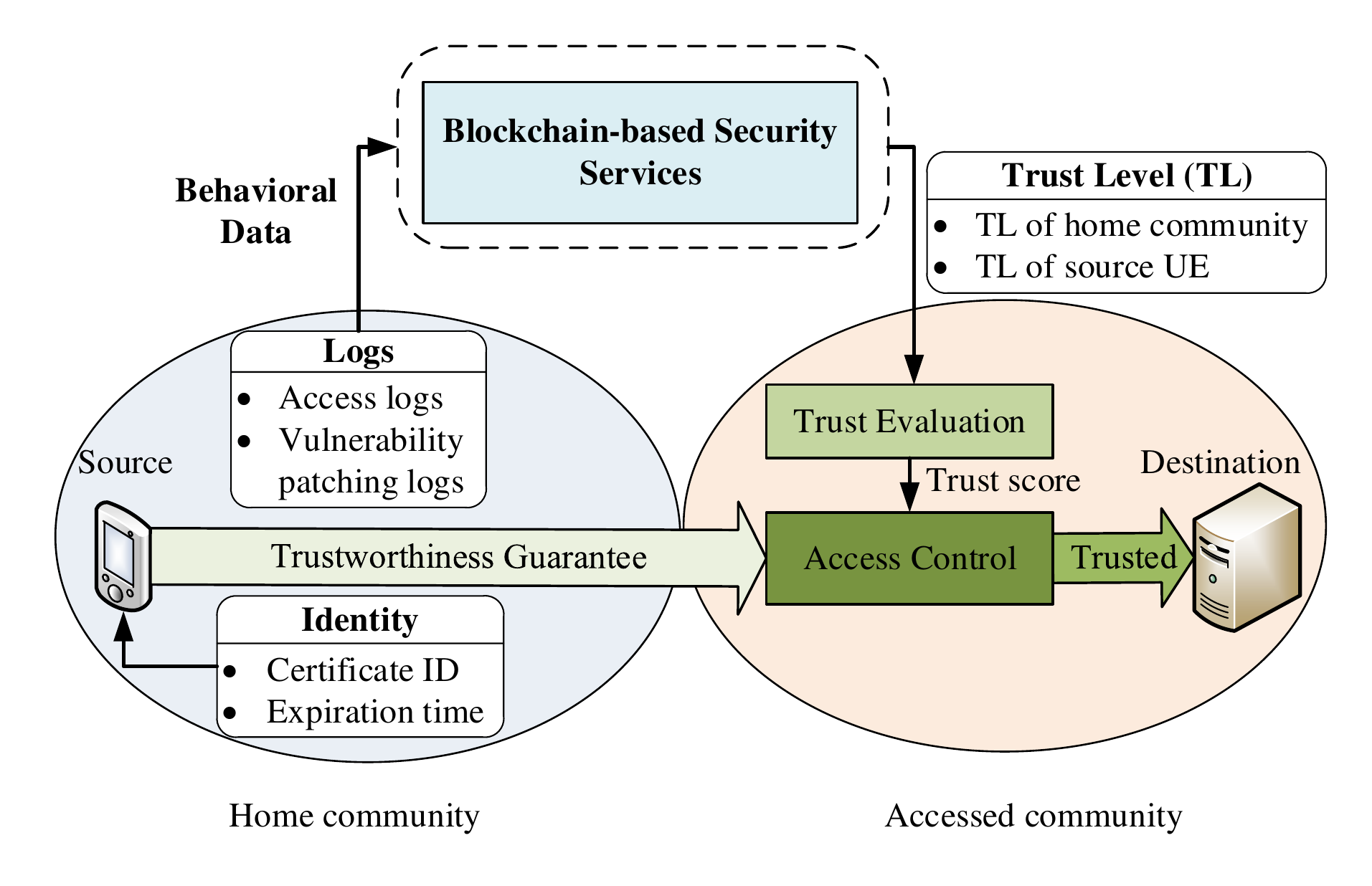}}
	\caption{The trust evaluation system in an access request.}
	\label{fig4}
\end{figure}
Trust evaluation is the process of quantifying trust values based on the attributes that affect trust. Trust in communication networks comes from the observation of the historical behaviors and recommendations of trusted third parties. In the proposed ZTA, the trust evaluation system works as shown in Fig. \ref{fig4}. The trustworthiness of a guest UE is preliminarily guaranteed by its home community through identity authentication and self-evaluation. In addition, the TBSSs implement trust-level assessments based on the CEL, vulnerability information, and behavioral data extracted from the home community. 

The information used for trust evaluation, including identity and behavioral data, is dependent on the trustworthiness of the home community. If the home community maliciously provides false UE identity information or forges logs, it seriously undermines the validity of the trust evaluation results and affects the security performance of the access control framework. Therefore, to ensure security performance, the trustworthiness evaluation of communities must be established as the trust root of the whole ZTA.

We introduce TPSSs as supervisors to evaluate the trustworthiness of communities. To this end, we consider the trust level of the home community as an important reference element in the trust evaluation of the guest UE. VDB can maintain a community vulnerability risk index through random probing to comprehensively evaluate the timeliness of vulnerability patching in the community, and impose indicator penalties on fraud and deception. CEL can employ techniques such as event traceability and network forensics to collect statistics on the distribution of attackers, virus sources, and bots to form a threat level evaluation index of communities. Similarly, ABD can perform statistical analysis on the frequency of abnormal access behaviors and other characteristics to generate a community behavior abnormality index. By incorporating these community trust-level indices as part of the trust assessment results of the guest UE, the approval probability of their access requests can be affected.

By introducing trust levels of communities as a reference in access control, community controllers can be effectively encouraged to focus on improving security performance to ensure the access authority of their customer UEs. Even though data fraud may not be completely eliminated, such behaviors can be effectively punished, thereby significantly reducing the occurrence probability of dishonest communities. Therefore, the foundation of the entire trust system can be consolidated to adapt to the openness of 6G.
\section{Numerical Results}
To verify the effectiveness of the proposed ZTA for 6G, that is, ZTA-6G, we compare it with two security architectures, Trust Based Packet Filtering (TBPF) \cite{TBPF} and Transparency for better Internet Security (TRIS) \cite{TRIS}. TBPF was developed to enable collaboration among multiple IDSs based on the trust evaluation of nodes and IP addresses. TRIS was proposed based on profound insights into DDoS attack defense, which can be applied to the security defense of the entire network. 

We set four communities, namely A, B, C and D, with 1,000 UEs each. We set links between A-B, A-C, B-D, and C-D. 
\begin{figure}[t]
	\centerline{\includegraphics[width=0.5\textwidth]{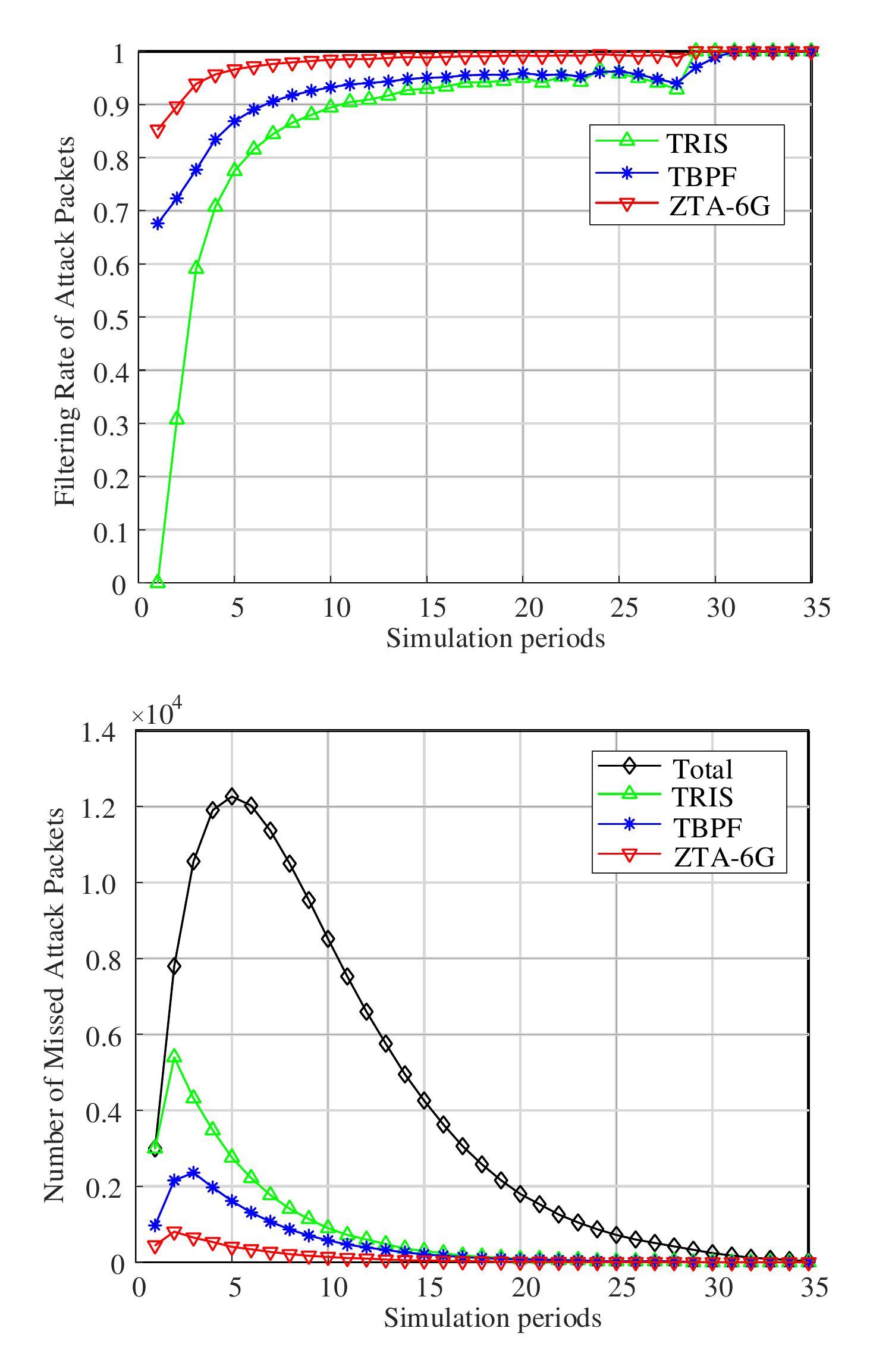}}
	\caption{Effectiveness of different security architectures in defending against non-spoofing DDoS attacks. The subfigure above illustrates the filtering rates of attack packets under different security architectures, and the subfigure below shows the number of missed attack packets.}
	\label{NSP}
\end{figure}

We adopt the susceptible-infectious-recovered (SIR) model \cite{SIR} to simulate the spread of worms in communities. Although remotely manipulating bots across communities can be prevented in ZTA-6G, bot-herders can still manipulate infected UEs within the same community to launch attacks. Multiple attackers can launch zero-day DDoS attacks through cross-community coordination. Before the attack is launched, we simulate a normal communication process lasting for 100 s with a cross-domain access probability of 0.1 at a rate of 5 packets per second (pps). The initial number of infected UEs is 100 per community. The infection and fixing rates of the vulnerability are both set to 0.2. Each period lasts for 1 second repeatedly until all vulnerable UEs are fixed. Because our framework is inherently resistant to IP spoofing, we evaluate the effectiveness of defense against non-spoofing attacks. The infected UEs in B and C first attack a victim in A, and then the infected UEs in A, B, and C jointly attack the victim in D. The attack intensity is set at 10 pps. We set the threshold of the trust value in TBPF to 0.75, as suggested by the authors \cite{TBPF}. For ZTA-6G, we adopt the same trust evaluation method as for TBPF. We collect the received attack packets at the victim side in community D to evaluate the defense performance of each architecture. The average results of 100 random simulations are shown in Fig. \ref{NSP}.
\begin{figure}[t]
	\centerline{\includegraphics[width=0.5\textwidth]{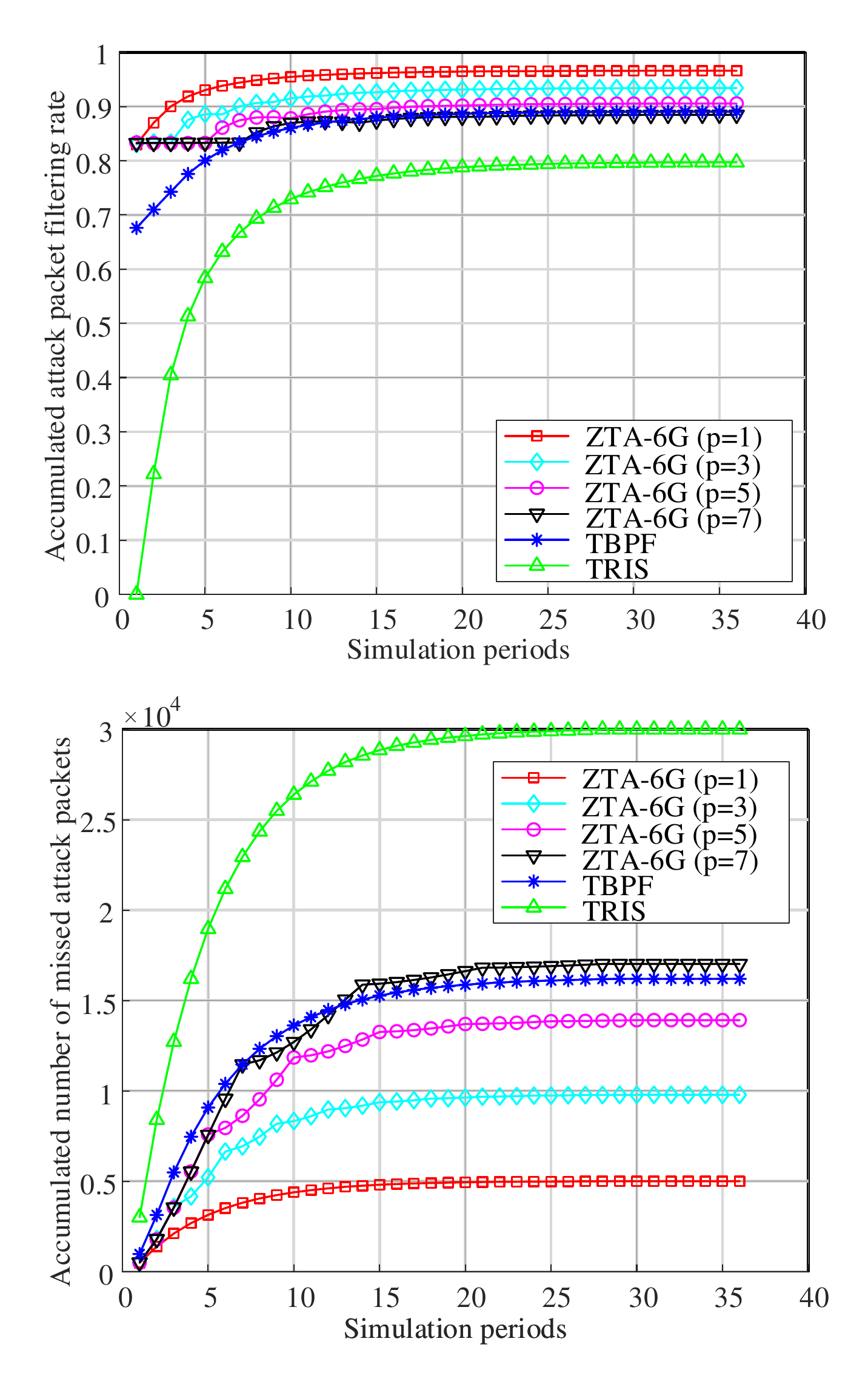}}
	\caption{Robustness of the proposed ZTA in different validity periods of the trust evaluation results. The subfigure above shows the accumulated filtering rates of attack packets by each method, and the subfigure below shows the accumulated number of missed attack packets.}
	\label{ValP}
\end{figure}

As shown in Fig. \ref{NSP}, as the simulation process proceeds, the defense effects of the three architectures generally improved, wherein ZTA-6G always performs the best. TRIS cannot defend against the first attack of malicious UEs, so its defense effect is the worst at the very beginning. The performance of TBPF is better. ZTA-6G filters out the packets from UEs with malicious behaviors in the source community by self-evaluation and re-evaluates the trustworthiness of vulnerable UEs in the destination community. Thus, it maintains the highest filtering rate of attack packets. Therefore, the number of missed attack packets is minimal (as shown in the subfigure below).

To verify the robustness of the ZTA-6G, we set different validity periods for the trust evaluation results instead of evaluating the trustworthiness of every access request. For the evaluation results of UEs within the validity period, ZTA-6G directly approves their access requests. Some UEs become bots owing to virus infection during this period, resulting in the degradation of security performance in terms of the accumulative filtering rate in non-spoofing DDoS attacks. We set the validity period to 1, 3, 5, and 7 s to verify the changes in the accumulated filtering rate of the attack packets and the number of missed attack packets by ZTA-6G. The average results of 100 random simulations are shown in Fig. \ref{ValP}.

As the validity period increases, the accumulative filtering rate of attack packets shows a downward trend, but the rate of decline slows down. Implementing trust evaluation every second ($p=1$) performs the best, and its time cost is the highest. For $p>1$, the accumulative filtering rate curves are wavy. This is because, after each re-evaluation, the effectiveness of the evaluation results declines over time. For $p=3$ and $p=5$, ZTA-6G performs much better than TRIS and TBPF. The accumulated detection rates are both higher than 90\%. Even if we re-evaluate the UE's trust value once every 7 s, the filtering rates of attack packets can still be much better than TRIS and are almost as good as TBPF. These results demonstrate the robustness of ZTA-6G. In addition, this also shows that it is feasible to sacrifice a small part of the security performance to obtain a significant improvement in time efficiency according to the safety requirements in applications.
\section{Open Issues}
To facilitate the implementation of ZTA-6G, there are still some open issues, such as system efficiency, mobility management, trust evaluation and residual security risks.
\subsection{System efficiency}
In ZTA-6G, the delay caused by the trust evaluation process may limit its application in delay-sensitive services. To reduce transmission latency, content distribution technologies can be used to deploy TPSSs so that communities can access their services nearby. The on-demand extraction of behavioral logs can be replaced by periodic active synchronization of each community, thereby reducing transmission delay. The behavior analysis algorithm can focus only on the correlation analysis of incremental data to improve processing speed. In addition, for UEs that frequently initiate access requests, TPSSs can set a validity period for the security assessment results by attaching a digital signature to the UE's identity certificate. The accessed community can verify the digital signature instead of initiating a new security service request. This mechanism can avoid repeated security assessments within the validity period, thereby reducing the response time for access authorization.
\subsection{Mobility management}
ZTA-6G stores and manages the identities, behavioral data and vulnerability fixing information of UEs based on communities. When a UE moves, its home community can change. The movement of network entities, such as satellites and drones, may also lead to changes in management domains. For delay-sensitive services, the handover management problem should be studied to ensure the timeliness of access control in the ZTA. A potential solution may arise from UE mobile pattern prediction-based artificial intelligence. Many studies have verified the predictability of human behavioral patterns \cite{predict}, and the regularity of the movement patterns of unmanned equipment may be more significant. Because data handovers caused by UE movement will only occur between adjacent communities, it suffices to perform data synchronization or predelivery for several communities that may be involved in the UE movement.
\subsection{Trust evaluation}
Trust evaluation methods face two major problems when applied to ZTA in 6G. In parameter determination, the weights of trust elements from different sources when calculating the trust value are usually determined by experience, such as the trust level of the community and UE. Moreover, the threshold of trust values often has difficulty balancing generalization and accuracy. There is no objective or quantitative parameter selection method. One possible solution is to develop data-driven trust evaluation methods based on federated learning techniques. Considering that it is difficult to collect sufficient behavioral data to train the learning models in a single community, the accuracy of models for the same type of UE will vary significantly in different communities. Establishing new model integration methods based on cross-community collaboration is required.
\subsection{Residual security risks}
Because trust evaluation requires time, attackers may manipulate bots to initiate a large number of access requests to the control plane of ZTA-6G in a specific community, resulting in DDoS attacks. Although ZTA-6G is resistant to cross-community manipulation and reuse of bots, an attacker could still control massive numbers of bots in the same community to launch such attacks. To solve this problem, a prescreening mechanism for access requests in the data plane of ZTA-6G is necessary. For example, prioritizing requests from repetitive visitors. Multiple existing studies on mitigating DDoS attacks against the SDN control plane can be referenced. In addition, ZTA-6G focuses mainly on cross-community security access and does not involve security issues within the community. Although the internal security problems of communities can be solved using existing methods, their integration with ZTA-6G requires further research. 

\section{Conclusions}
To satisfy the new security requirements of 6G, we have designed a software-defined ZTA to collaboratively defend against network threats. In the new architecture, the community can implement sophisticated access control to guest UEs with the help of trust assessment from TPSSs. Distributed identity management is achieved through a newly designed digital certificate system. The trust system of this architecture is established by employing TPSSs as supervisors to solidify the trust roots. We have verified the effectiveness and robustness of the proposed ZTA by simulating worm-spreading and zero-day DDoS attacks.
\ifCLASSOPTIONcaptionsoff
  \newpage
\fi

\bibliographystyle{IEEEtran}
\bibliography{ZTNbib}

\begin{thebibliography}{10}
\providecommand{\url}[1]{#1}
\csname url@samestyle\endcsname
\providecommand{\newblock}{\relax}
\providecommand{\bibinfo}[2]{#2}
\providecommand{\BIBentrySTDinterwordspacing}{\spaceskip=0pt\relax}
\providecommand{\BIBentryALTinterwordstretchfactor}{4}
\providecommand{\BIBentryALTinterwordspacing}{\spaceskip=\fontdimen2\font plus
\BIBentryALTinterwordstretchfactor\fontdimen3\font minus
  \fontdimen4\font\relax}
\providecommand{\BIBforeignlanguage}[2]{{%
\expandafter\ifx\csname l@#1\endcsname\relax
\typeout{** WARNING: IEEEtran.bst: No hyphenation pattern has been}%
\typeout{** loaded for the language `#1'. Using the pattern for}%
\typeout{** the default language instead.}%
\else
\language=\csname l@#1\endcsname
\fi
#2}}
\providecommand{\BIBdecl}{\relax}
\BIBdecl

\bibitem{Fang}
X.~Fang, W.~Feng, T.~Wei, Y.~Chen, N.~Ge, and C.-X. Wang, ``5g embraces
  satellites for 6g ubiquitous iot: basic models for integrated satellite
  terrestrial networks,'' \emph{IEEE Internet of Things Journal}, vol.~8,
  no.~18, pp. 14\,399--14\,417, 2021.

\bibitem{csm}
D.~Je, J.~Jung, and S.~Choi, ``Toward 6g security: Technology trends, threats,
  and solutions,'' \emph{IEEE Communications Standards Magazine}, vol.~5,
  no.~3, pp. 64--71, 2021.

\bibitem{survey}
V.-L. Nguyen, P.-C. Lin, B.-C. Cheng, R.-H. Hwang, and Y.-D. Lin, ``Security
  and privacy for 6g: A survey on prospective technologies and challenges,''
  \emph{IEEE Communications Surveys \& Tutorials}, vol.~23, no.~4, pp.
  2384--2428, 2021.

\bibitem{BeyondZT}
M.~{Campbell}, ``{Beyond Zero Trust: Trust Is a Vulnerability},''
  \emph{Computer}, vol.~53, no.~10, pp. 110--113, 2020.

\bibitem{book2017}
E.~Gilman and D.~Barth, \emph{Zero Trust Networks}, 1st~ed.\hskip 1em plus
  0.5em minus 0.4em\relax Sebastopol, CA: O’Reilly Media, Inc., July 2017.

\bibitem{BeyondCorp}
R.~{Ward} and B.~{Beyer}, ``Beyondcorp: A new approach to enterprise
  security,'' \emph{;login:}, vol.~39, no.~6, pp. 6--11, 2014.

\bibitem{progress2020}
\BIBentryALTinterwordspacing
H.~Schulze, ``\BIBforeignlanguage{en}{{Zero Trust Progress Report}},''
  Cybersecurity Insiders, Tech. Rep., 2020. [Online]. Available:
  \url{https://www.cybersecurity-insiders.com/portfolio/2020-zero-trust-progress-report-pulse-secure/}
\BIBentrySTDinterwordspacing

\bibitem{NIST}
\BIBentryALTinterwordspacing
S.~Rose, O.~Borchert, S.~Mitchell, and S.~Connelly,
  ``\BIBforeignlanguage{en}{{Zero Trust Architecture}},'' National Institute of
  Standards and Technology, Tech. Rep., 2020. [Online]. Available:
  \url{https://doi.org/10.6028/NIST.SP.800-207}
\BIBentrySTDinterwordspacing

\bibitem{SDP}
A.~Moubayed, A.~Refaey, and A.~Shami, ``{Software-Defined Perimeter (SDP):
  State of the Art Secure Solution for Modern Networks},'' \emph{IEEE Network},
  vol.~33, no.~5, pp. 226--233, 2019.

\bibitem{You}
X.~You, C.~Wang, and J.~Huang, ``Towards 6g wireless communication networks:
  vision, enabling technologies, and new paradigm shifts,'' \emph{Science China
  Information Sciences}, vol.~64, no. 11301, pp. 1--74, 2021.

\bibitem{DT}
Y.~Wu, K.~Zhang, and Y.~Zhang, ``Digital twin networks: A survey,'' \emph{IEEE
  Internet of Things Journal}, vol.~8, no.~18, pp. 13\,789--13\,804, 2021.

\bibitem{TBPF}
W.~Meng, W.~Li, and L.~F. Kwok, ``Towards effective trust-based packet
  filtering in collaborative network environments,'' \emph{IEEE Transactions on
  Network and Service Management}, vol.~14, no.~1, pp. 233--245, 2017.

\bibitem{TRIS}
C.~Pappas, T.~Lee, R.~M. Reischuk, P.~Szalachowski, and A.~Perrig, ``Network
  transparency for better internet security,'' \emph{IEEE/ACM Transactions on
  Networking}, vol.~27, no.~5, pp. 2028--2042, 2019.

\bibitem{SIR}
W.~K. Chai, ``Modelling spreading process induced by agent mobility in complex
  networks,'' \emph{IEEE Transactions on Network Science and Engineering},
  vol.~5, no.~4, pp. 336--349, 2018.

\bibitem{predict}
F.~Xia, J.~Wang, X.~Kong, Z.~Wang, J.~Li, and C.~Liu, ``Exploring human
  mobility patterns in urban scenarios: A trajectory data perspective,''
  \emph{IEEE Communications Magazine}, vol.~56, no.~3, pp. 142--149, 2018.

\end{thebibliography}
\end{document}